# Role of valence changes and nanoscale atomic displacements in BiS$_2$-based superconductors

Jie Cheng[1*], Huifei Zhai[2], Yu Wang[3], Wei Xu[4,5], Shengli Liu[1,6] and Guanghan Cao[2]

[1]Center of Advanced Functional Ceramics, College of Science, Nanjing University of Posts and Telecommunications, Nanjing, Jiangsu 210023, China [2]Department of Physics, Zhejiang University, Hangzhou 310027, China [3]Shanghai Synchrotron Radiation Facility, Shanghai Institute of Applied Physics, Chinese Academy of Sciences, Shanghai 201204, China [4]Beijing Synchrotron Radiation Facility, Institute of High Energy Physics, Chinese Academy of Sciences, Beijing 100049, China [5]Rome International Center for Materials Science, Superstripes, RICMASS, via dei Sabelli 119A, I-00185 Roma, Italy [6]Nanjing University (Suzhou) High-Tech Institute, Suzhou, 215123, China (email:chengj@njupt.edu.cn)

**Superconductivity within layered crystal structures has attracted sustained interest among condensed matter community, primarily due to their exotic superconducting properties. EuBiS$_2$F is a newly discovered member in the BiS$_2$-based superconducting family, which shows superconductivity at 0.3 K without extrinsic doping. With 50 at.% Ce substitution for Eu, superconductivity is enhanced with Tc increased up to 2.2 K. However, the mechanisms for the $T_c$ enhancement have not yet been elucidated. In this study, the Ce-doping effect on the self-electron-doped superconductor EuBiS$_2$F was investigated by X-ray absorption spectroscopy (XAS). We have established a relationship between Ce-doping and the $T_c$ enhancement in terms of Eu valence changes and nanoscale atomic displacements. The new finding sheds light on the interplay among superconductivity, charge and local structure in BiS$_2$-based superconductors.**

Superconductivity in quasi-two-dimensional crystal structures has attracted sustained interest in the past decades. The most outstanding examples include high-$T_c$ cuprates with CuO$_2$ superconducting layers[1] and Fe-based superconductors with a Fe-square lattice[2]. Very recently, superconductivity of BiS$_2$-based compounds which have similar layered crystal structure as those of cuprates and Fe-based materials has been reported. The first member of the BiS$_2$-based superconducting family is Bi$_4$O$_4$S$_3$ with a $T_c$ of 8.6 K[3]. It was found that the characteristic BiS$_2$ layers are responsible for the superconductivity[3].

So far, several ReBiS$_2$O$_{1-x}$F$_x$ (Re = La, Ce, Pr and Nd) and doped SrBiS$_2$F superconductors have been discovered with the highest $T_c$ of 10.6 K[4-10]. Band structure calculations indicate that the undoped parent compounds such as LaBiS$_2$O and SrBiS$_2$F are insulators with an energy gap of 0.82 and 0.80 eV, respectively[11,12]. Upon electron doping, both compounds exhibit metallic conducting behavior and a superconducting transition at low temperatures[4,10]. On the other hand, recent works demonstrate that the isostructural compounds EuBiS$_2$F and Eu$_3$Bi$_2$S$_4$F$_4$ are metallic, and they even exhibit superconductivity without extrinsic doping, at temperatures below 0.3 K and 1.5 K respectively[13,14], different from the other analogues. By various experimental approaches, it is pointed out that the self-doping nature of the observed superconductivity in both EuBiS$_2$F and Eu$_3$Bi$_2$S$_4$F$_4$ is due to the mixed valence of Eu[13,14]. Currently, with 50 at.% Ce substitution for Eu in EuBiS$_2$F, the $T_c$ is enhanced up to 2.2 K[15]. It was suggested that the Eu valence is essentially divalent in Ce-doped system[15]. On the contrary, the average Eu valence with respect to the parent compound increases with the Se doping in Eu$_3$Bi$_2$S$_{4-x}$Se$_x$F$_4$ which has the highest $T_c$ of 3.35 K[16]. How the Eu valence changes and its consequence on superconductivity in the parent and doped BiS$_2$-based superconductors still remain unresolved.

Moreover, one of the important problems in the layered systems is the inter- and intra-layer interactions. Similar to Fe-based superconductors, the interactions between superconducting BiS$_2$ layers and blocking layers can be revealed *via* the nanoscale atomic displacements[17,18]. Hence, in order to understand the origin of superconductivity, it is critical to investigate the Eu valence and the local atomic displacements in the parent and doped Eu-containing BiS$_2$-based superconductors.

The X-ray absorption spectroscopy (XAS), consisting of the X-ray absorption near edge spectroscopy (XANES) and extended X-ray absorption fine structure (EXAFS) spectroscopy, is an ideal technique to retrieve the substantial information of both valence transition and nanoscale atomic displacements, thus XAS has been widely applied in physics and chemistry[19-21]. For example, based on the "fingerprint effect", Eu $L_3$-edge XANES for EuFe$_2$As$_2$ presents the visually experimental evidence for the pressure-induced valence changes of Eu ions[22]. In addition, Bi $L_3$-edge EXAFS were performed to probe the local atomic structure of BiS$_2$-based systems[18]. In this contribution, we investigated the local structure of EuBiS$_2$F-based system as a function of Ce-doping by XAS, providing the atomic site-selective information of valence changes and nanoscale atomic displacements.

**Results**

**Role of Eu valence changes in the parent and Ce-doped EuBiS$_2$F.** For the Eu-containing superconductors, detailed investigations of the Eu valence change may provide valuable information on the electronic structure, which is fundamental for a better understanding of their superconductivity[22,23]. Figure 1a shows normalized Eu $L_3$-edge XANES data for EuBiS$_2$F and Eu$_{0.5}$Ce$_{0.5}$BiS$_2$F. The main peak (6975 eV) and the other feature (6983 eV) in the Fig. 1a are associated respectively to Eu$^{2+}$ ($4f^{\,7}$) and Eu$^{3+}$ ($4f^{\,6}$)[22].

Now we determine quantitatively the valence of Eu for the parent and Ce-doped EuBiS$_2$F by fitting the XANES spectra to an arctangent step function and a Lorentzian

peak for each valence state. The mean valence was determined by using a widely used method[24,25]:

$$v = 2 + [I^{3+}/(I^{2+} + I^{3+})] \qquad (1)$$

where $I^{2+}$ and $I^{3+}$ is integrated intensity of peaks corresponding to $Eu^{2+}$ and $Eu^{3+}$ on XANES spectrum. Based on the best curve fit in Fig. 1, we estimated the mean valence of Eu ions in $EuBiS_2F$ is +2.16(1), instead of +2, demonstrating the self-electron-doping nature in parent compound without any extrinsic doping. The mean valence of Eu in Ce-doped $EuBiS_2F$ is +2.05(1), basically consistent with previous crystallographic and magnetic structure data[15]. Therefore, these data confirm the Eu valence change, suggesting a potential relationship between the Eu valence and superconductivity.

In Fig. 2 we focus on the normalized Ce $L_3$-edge XANES in $Eu_{0.5}Ce_{0.5}BiS_2F$, in which three main structures A, B and C can be identified. The first peak A around 5728 eV is associated to the transition from the Ce 2$p$ core level to the vacant Ce 5$d$ state mixed with the Ce 4$f^1$ final state, i.e. $Ce^{3+}$ state[26]. On the other hand, the weak feature B around 5745 eV is a characteristic feature of layered rare-earth systems[26], and its intensity is generally sensitive to the F atom order/disorder in the Eu/CeF layers. The third peak C is the so-called continuum resonance, providing the information on the local lattice structures. It should be noted that the energy difference between the characteristic $Ce^{3+}$ (4$f^1$) and $Ce^{4+}$ (4$f^0$) absorption peaks is approximately 12 eV, which is independent and is mainly determined by the Ce 2$p$-4$f$ Coulomb interaction[26]. But in Fig. 2 we found no obvious evidence of $Ce^{4+}$ feature around 5740 eV, demonstrating that the Ce valence in the $Eu_{0.5}Ce_{0.5}BiS_2F$ sample is essentially trivalent. Considering the valence of Eu, 50 at.% Ce-doping could cause an increment of mean valence for Eu/Ce ions, which increases from +2.16 of parent $EuBiS_2F$ to +2.53 of Ce-doped system. Consequently, additional 17% charges were induced upon Ce-doping in $EuBiS_2F$, which is believed to be crucial for the superconductivity enhancement.

**Nanoscale atomic displacements in $EuBiS_2F$ and $Eu_{0.5}Ce_{0.5}BiS_2F$.** As is well known, material properties are in a close relationship with its nanoscale atomic structure. Analogous to cuprates and Fe-based superconductors, Ce impurity could alter the local atomic displacements of both blocking layers and $BiS_2$ superconducting layers. Therefore, to gain an insight into the atomic displacements induced by Ce-doping, we have undertaken detailed structural study by means of Eu and Bi $L_3$-edge EXAFS measurements. Figure 3 and 4 display the Fourier transform (FT) magnitudes of the EXAFS oscillations providing real space information at Eu and Bi $L_3$-edge, respectively. We have to underline that the positions of the peaks in the FT are shifted a few tenths of Å from the actual interatomic distances because of the EXAFS phase shift[27]. In the $BiS_2$ layer the in-plane and out-of-plane S atoms are denoted as S1 and S2, respectively. The Eu atom is coordinated with four nearest F atoms at ~ 2.52 Å and four S2 atoms at ~ 3.04 Å. Therefore, the broad structure ($R$ = 1.5 ~ 3.0 Å) in the FT of Eu $L_3$-edge EXAFS corresponds to the contributions of Eu-F and Eu-S2 bonds. On the other hand, the near-neighbor of Bi atoms are one out-of-plane S2 atom at ~ 2.50 Å and four in-plane S1 atoms at ~ 2.87 Å. Therefore, the broad structure ($R$ = 1.4 ~ 2.6 Å) in Fig. 4 contains information on the Bi-S2 and Bi-S1 bonds. Obviously, large changes in the FTs of both Eu and Bi $L_3$-edge can be seen with Ce-doping, indicating the atomic displacements in blocking layers and also in the electronically active BiS$_2$ layers.

The EXAFS amplitude depends on several factors and is given by the following general equation [28]:

$$\chi(k) = \sum_j \frac{N_j S_0^2}{k R_j^2} f_j(k, R_j) \exp[-2k^2 \sigma_j^2] \exp[\frac{-2R_j}{\lambda}] \sin[2kR_j + \delta_j(k)] \qquad (2)$$

where $N_j$ is the number of neighboring atoms at a distance $R_j$, $S_0^2$ is the passive electron reduction factor, $f_j(k, R_j)$ is the backscattering amplitude, $\lambda$ is the photoelectron mean free path, $\delta_j(k)$ is the phase shift and $\sigma_j^2$ is the correlated Debye-Waller factor.

In order to obtain quantitative results, we firstly fit the peaks of EXAFS spectra at Eu $L_3$-edge involving contributions of four Eu-F and four Eu-S2 bonds, which were isolated from the FTs with a rectangular window. The range in $k$ space was 3 ~ 12 Å$^{-1}$ and that in $R$ space was 1.5 ~ 3.0 Å. Considering the absorption energy at Eu $L_3$ (6977 eV) and $L_2$-edge (7617 eV), the maximum wave-vector $k$ for Eu $L_3$-edge EXAFS is up to 12 Å$^{-1}$. The spatial resolution $\Delta R = \pi/2k_{max}$[28] is about 0.13 Å with the $k_{max} = 12$ Å$^{-1}$, which is sufficient to distinguish between Eu-F and Eu-S2 bonds. For the least-squares fits, average structure measured by diffraction on EuBiS$_2$F system[13] is used as the starting model. The backscattering amplitudes and phase shift were calculated using the FEFF code[29]. Only the radial distances $R_j$ and the corresponding $\sigma_j^2$ were allowed to vary, with coordination numbers $N_j$ fixed to the nominal values. The passive electrons reduction factor $S_0^2$ and photoelectron energy zero $E_0$ were also fixed after fit trials on different scans. The best values for the $S_0^2$ were found to be 0.9 and fixed to this value for all the shells. The number of independent parameters which could be determined by

EXAFS is limited by the number of the independent data points $N_{ind} \sim (2\Delta k \Delta R)/\pi$, where $\Delta k$ and $\Delta R$ are respectively the ranges of the fit in the $k$ and $R$ space[28]. In our case, $N_{ind}$ is 8 ($\Delta k$ = 9 Å$^{-1}$, $\Delta R$ = 1.5 Å), sufficient to obtain all parameters.

As shown in Table 1, upon Ce-doping the distance of Eu-S2 bond is essentially unchanged within the errors, while the Eu-F distance becomes slightly elongated from 2.51(1) Å to 2.54(1) Å, suggesting a thicker EuF layer induced by Ce-doping. Now we resort to the bond valence sum[30] of Eu (Eu-BVS) using the formula $\sum \exp(\frac{R_0 - d_{ij}}{0.37})$, where $R_0$ is an empirical parameter (2.04 and 2.53 Å for Eu-F and Eu-S bonds[30], respectively) and $d_{ij}$ denotes the measured bond distances between Eu and coordinate anions. Here, eight coordinate atoms (four F and four S2 atoms) were considered. Considering the bondlengths achieved from EXAFS fitting, the Eu-BVS value are +2.14(2) and +2.07(2) in EuBiS$_2$F and Eu$_{0.5}$Ce$_{0.5}$BiS$_2$F respectively, essentially in agreement with the valence information retrieved from our XANES data.

Meanwhile, Ce-doping also affects the local atomic structure of superconducting BiS$_2$ layers. In Fig. 4 the broad peaks at Bi $L_3$-edge were modelled by two shells, involving contributions of one Bi-S2 and four Bi-S1 bonds, which were isolated from the FTs with a rectangular window. The range in $k$ space was 3 ~ 15 Å$^{-1}$ and that in $R$ space was 1.4 ~ 2.6 Å. Spatial resolution $\Delta R = \pi/2k_{max}$ is about 0.10 Å, while the number of independent parameters $N_{ind}$ is 9, sufficient to distinguish between Bi-S2 and Bi-S1 bonds and obtain all parameters.

Recently, it was reported that the enhancement of in-plane chemical pressure is responsible for the superconductivity in BiS$_2$-based compounds[31]. Upon Ce-doping the sharp contraction of the in-plane Bi-S1 bond ($\Delta R \sim 0.11$Å, *i.e.* a higher in-plane chemical pressure) results in an enhancement of the packing density of Bi and S1 ions within the superconducting plane, which would enhance the hybridization of Bi $6p_x/6p_y$-S $3p$ orbitals and result in an increase of $T_c$. In addition, the fact that in-plane Bi-S1 bondlength decreases with Ce-doping, while the Bi-Bi distance (*i.e. a*-axis, from 4.0508(1) to 4.0697(1) Å) showing a small increase, indicating the puckering and large in-plane disorder of the Bi-S1 layer. Further information on the atomic disorder can be provided by the correlated Debye-Waller factors ($\sigma^2$), measuring the mean square relative displacement (MSRD) of the photoabsorber-backscatterer pairs[32]. Data point out that the $\sigma^2$ for the in-plane Bi-S1 distance in EuBiS$_2$F is anomalously large, demonstrating a large configurational disorder within the Bi-S1 plane. Here, it is worth recalling that the large configurational disorder in BiS$_2$ plane is quite common in BiS$_2$-based superconductors, consistent with the anomalously large diffraction thermal factor of in-plane S1 atom[33]. Upon Ce-doping, the $\sigma^2$ for the Bi-S1 bond reduces by 25% with respect to the parent compound, demonstrating that puckering of the Bi-S1 layer seems to be getting reduced; that is to say, a flatter Bi-S1 plane is also responsible for a higher $T_c$. By contrast, the $\sigma^2$ for the Bi-S2 bond is quite small and remains unchanged upon Ce-doping, indicating robust Bi $6p_z$-S $3p$ hybridizations. All these results suggest that Ce-doping can effectively tune the atomic displacements of BiS$_2$ superconducting layers.

**Discussion**

The Ce-doping effect on the valence state and local atomic displacement in the EuBiS$_2$F

system is investigated by using XAS measurements. First of all, the valence of Eu ions in EuBiS$_2$F is estimated to be about +2.16(1), demonstrating the self-electron-doping nature without any extrinsic doping. Upon 50 at.% Ce-doping, the mean valence of Eu reduces to +2.05(1) and that of Ce ions are essentially trivalent. The main effect of Ce-doping is to provide additional 17% electrons into the system, beneficial for the superconductivity enhancement. The local atomic displacements can be revealed by Eu and Bi $L_3$-edge EXAFS: 1) the in-plane Bi-S1 distance is characterized by a large configurational disorder in EuBiS$_2$F-based system, which is quite common in BiS$_2$-based superconductors; 2) both the shortening of the in-plane Bi-S1 bond (*i.e.* a higher in-plane chemical pressure) and the flatter Bi-S1 plane are responsible for an enhancement of superconductivity.

In summary, we established a relationship between Ce-doping and the $T_c$ enhancement in EuBiS$_2$F-based superconductors, in terms of valence changes and nanoscale atomic displacements. The new findings are promising for providing insights on the interplay of charge, local structure and superconductivity.

**Methods**

Polycrystalline compounds of EuBiS$_2$F and Eu$_{0.5}$Ce$_{0.5}$BiS$_2$F were synthesized by solid-state reaction method[13,15]. The samples were well characterized for their phase purity, superconducting and other properties prior to the XAS measurements. The XAS spectra were collected at the BL-14W1 beamline of Shanghai Synchrotron Radiation Facility (SSRF). The storage ring was working at electron energy of 3.5 GeV, and the maximum stored current was about 250mA. The energy of the incident energy was tuned by scanning a Si (111) double crystal monochromator with energy resolution about 10$^{-4}$. The XAS spectra at Ce $L_3$-edge, Eu $L_3$-edge, and Bi $L_3$-edge were collected with several scans in transmission mode at room temperature. Data reduction was performed using the IFEFFIT program package[34].

**Acknowledgement**

This work was partly supported by the National Natural Science Foundation of China (NSFC 11405089 and U1532128), the Natural Science Foundation of Jiangsu Province of China (No. BK20130855), the "Six Talents Peak" Foundation of Jiangsu Province (2014-XCL-015), the Nanotechnology Foundation of Suzhou Bureau of Science and Technology (ZXG201444) and the Scientific Research Foundation of Nanjing University of Posts and Telecommunications (No. NY213053).

**Author contribution statement**

J.C. performed the experiment and analyzed the data. Y. W. and W. X. provided the support for the data collection and analysis. G. H. C and H. F. Z. provided the samples and discussed the results. J.C. and S. L. L. wrote the paper. All of the authors reviewed on the manuscript.

**Additional Information**

**Competing financial interests:** The authors declare no competing financial interests.

**Figure caption**

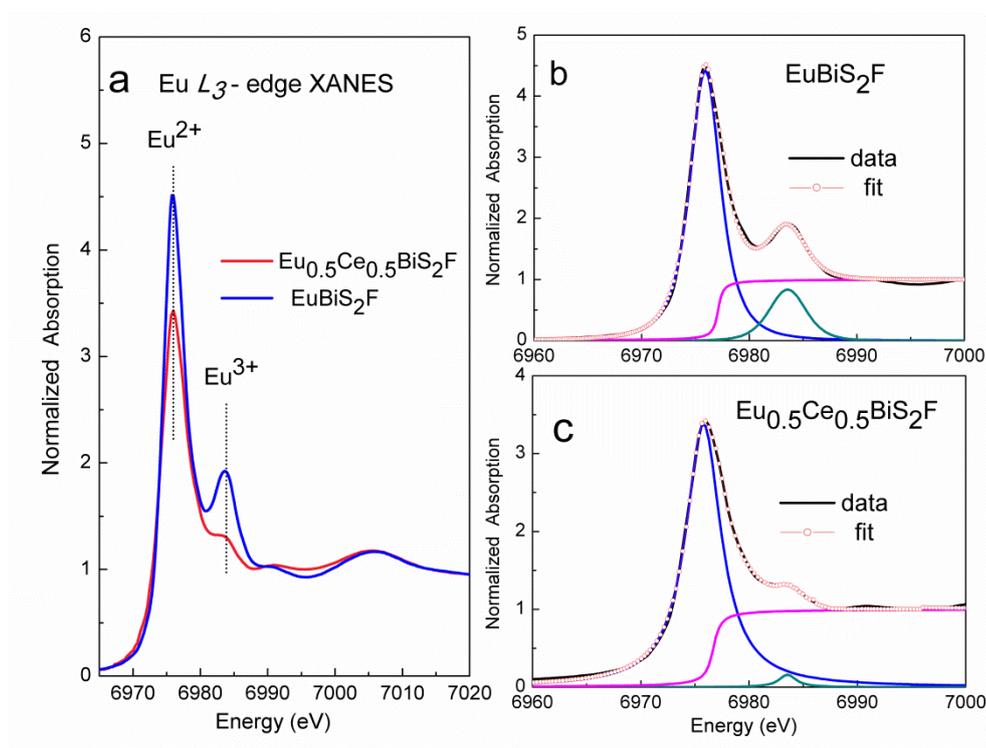

Figure 1. Eu $L_3$-edge XANES spectra and curve fitting for $EuBiS_2F$ and $Eu_{0.5}Ce_{0.5}BiS_2F$. (a) Normalized Eu $L_3$-edge XANES spectra for $EuBiS_2F$ and $Eu_{0.5}Ce_{0.5}BiS_2F$; (b) curve fitting for $EuBiS_2F$; (c) curve fitting for $Eu_{0.5}Ce_{0.5}BiS_2F$. The solid black line and red open circles correspond to the experimental data and the best fit, respectively.

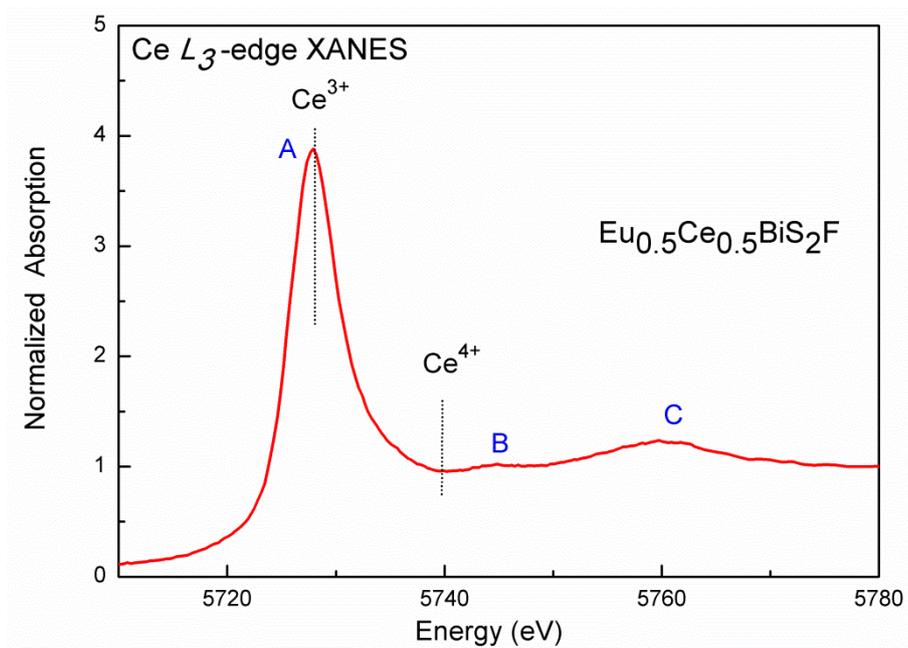

Figure 2. Normalized Ce $L_3$-edge XANES data for $Eu_{0.5}Ce_{0.5}BiS_2F$.

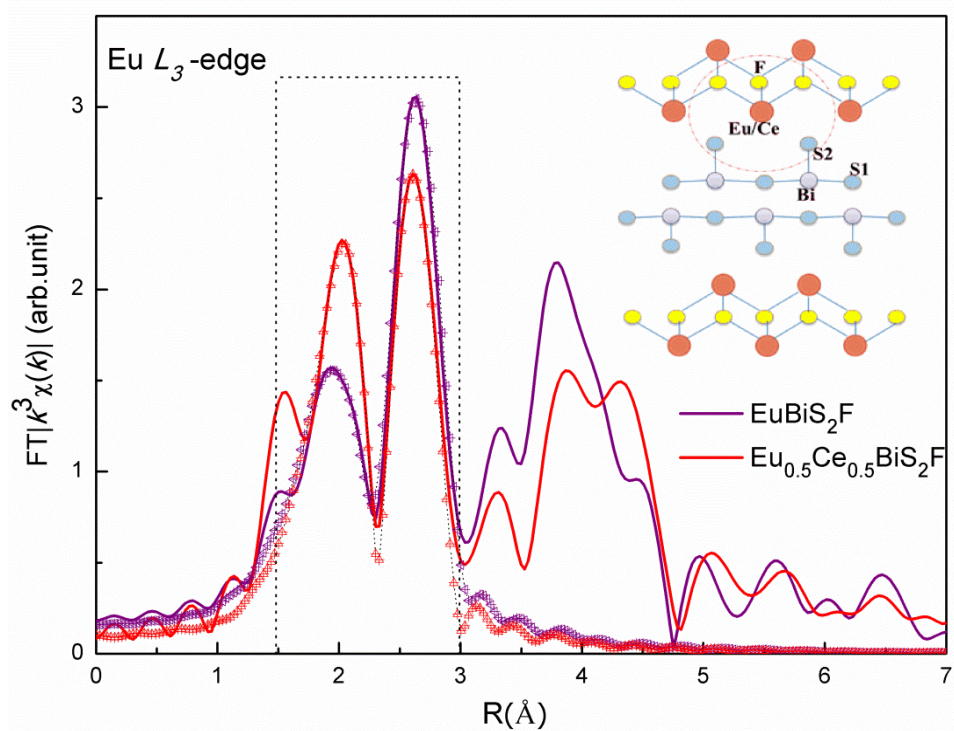

Figure 3. Fourier transform (FT) magnitudes of the Eu $L_3$-edge EXAFS measured on $EuBiS_2F$ and $Eu_{0.5}Ce_{0.5}BiS_2F$. Models fits to the FTs are also shown as triangles. The inset shows the local coordinate atomic clusters around Eu in cross-section view.

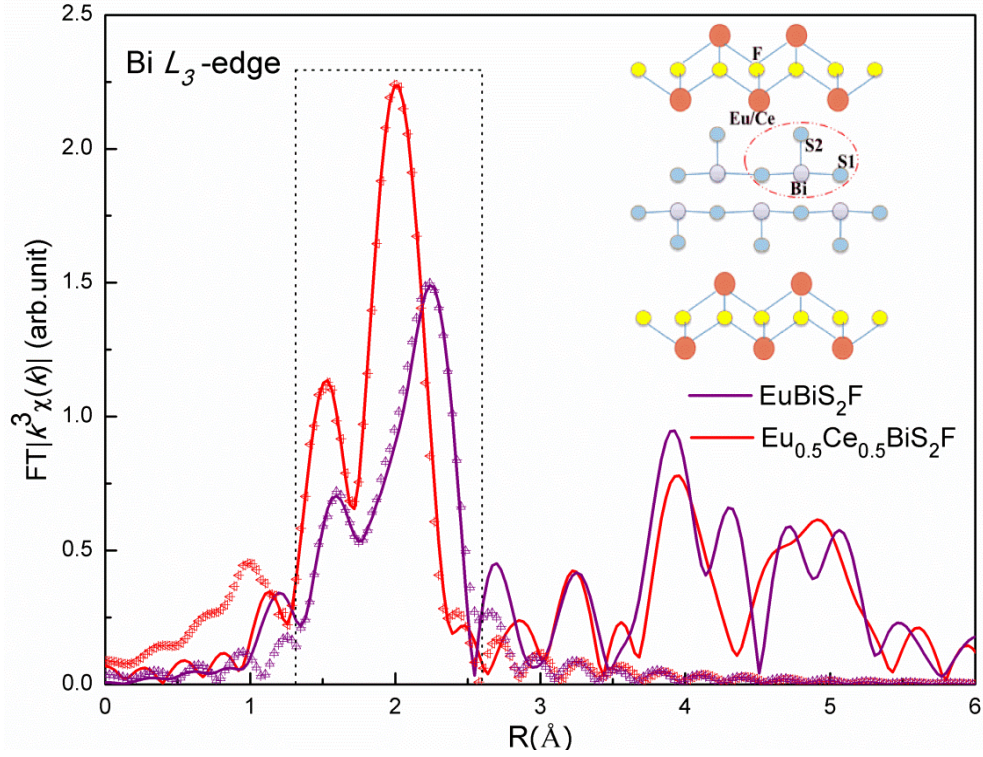

Figure 4. Fourier transform (FT) magnitudes of the Bi $L_3$-edge EXAFS measured on EuBiS$_2$F and Eu$_{0.5}$Ce$_{0.5}$BiS$_2$F. Models fits to the FTs are also shown as triangles. The inset shows the local coordinate atomic clusters encircled around Bi in cross-section view.

| System | | EuBiS$_2$F | Eu$_{0.5}$Ce$_{0.5}$BiS$_2$F |
|---|---|---|---|
| Eu-F | $R$ (Å) | 2.51(1) | 2.54(1) |
| | $\sigma^2$ ($10^{-3}$ Å$^2$) | 15.1(1) | 12.5(2) |
| Eu-S2 | $R$ (Å) | 3.04(2) | 3.03(1) |
| | $\sigma^2$ ($10^{-3}$ Å$^2$) | 10.3(2) | 12.6(1) |
| Bi-S2 | $R$ (Å) | 2.49(2) | 2.48(1) |
| | $\sigma^2$ ($10^{-3}$ Å$^2$) | 2.5(3) | 2.6(1) |
| Bi-S1 | $R$ (Å) | 2.79(2) | 2.68(1) |
| | $\sigma^2$ ($10^{-3}$ Å$^2$) | 31.2(2) | 23.4(1) |

Table 1 The fitting result at Eu and Bi $L_3$-edge EXAFS upon Ce-doping. The errors represent maximum uncertainty, determined using correlation maps between different parameters and by analysing different EXAFS scans.